\documentclass{PoS}

\title{Heavy flavour production at LHC}

\ShortTitle{Heavy flavour production at LHC}

\author{\speaker{Alessandro GRELLI}\thanks{The European Research Council has provided financial support under the European Community's Seventh Framework Programme (FP7/2007-2013) / ERC grant agreement no 210223. This work was supported in part by a Vidi grant from the Netherlands Organization for Scientific Research (project number: 680-47-232).} and Andr\'e MISCHKE$^{\dagger}$.\\ 
ERC-Starting Independent Research Group, Faculty of Science, Utrecht University, Princetonplein 5, 3584CC Utrecht, The Netherlands.\\ 
E-mail: \email{a.grelli@uu.nl, a.mischke@uu.nl}}

\abstract{
The Large Hadron Collider (LHC) will open a new era in high energy physics. The expected large cross section for heavy flavour production in proton-proton collisions at $\sqrt{s}$ = 14 TeV will allow detailed studies of the production mechanisms and an extensive test of Quantum Chromodynamics. Since charm and beauty has been proposed as a good probe to study hot and dense QCD matter, the understanding of the production mechanisms in elementary proton-proton collisions is of primary importance as a reference for studies in heavy-ion collisions. In the early phase of LHC operation the experiments will focus on the investigation of the heavy flavour production mechanisms.
}

\FullConference{The 2009 Europhysics Conference on High Energy Physics,\\
July 16 - 22 2009\\
Krakow, Poland}

\begin{document}

\section{Introduction}

Since more than 30 years heavy flavour production at hadronic colliders is one of the main fields of investigation in high energy and heavy-ion physics. In the last few years,  thanks to a strong theoretical effort, great improvements are achieved in the understanding of $c$ and $b$ production at the Tevatron. 
On one side the \textit{b} production cross-section is well described by theory expectations \cite{cdffirst}, on the other side the data points for charmed mesons cross sections are on the edge of theoretical errors and the low $p_T$ part of the spectra is not accessible \cite{cdfsecond}. In the last decade results from the Relativistic Heavy Ion Collider (RHIC) experiments at Brookeanven National Laboratory became available. In particular, the STAR and PHENIX experiments measured the charm production cross-section at $\sqrt{s}$ = 200 GeV in proton-proton, deuton-gold and gold-gold collisions. Although a factor two discrepancy in the central values, the measurements are in agreement with the model calculations within errors. 

In this paper we introduce the heavy flavour production at LHC reviewing the ALICE~\cite{ALICErep}, ATLAS~\cite{ATLASrep} and CMS~\cite{CMSrep} programs with early data. The LHCb program is not covered in this paper, for references see~\cite{lhcb, lhcb2, lhcb3}. For CMS and ATLAS we will focus on the possible first physics results using quarkonia states. In the CMS case the inclusive \textit{b} production will be introduced. The ALICE detector, considering the excellent inner tracker, the low material budget and the 0.5 T magnetic field is expected to allow, in addition to quarkonia measurements, extensive open heavy flavour studies both in proton-proton (p-p) and nucleus-nucleus (A-A) collisions.  

\section{Heavy flavour production at LHC}

The Large Hadron Collider, thanks to the center of mass energy of $\sqrt{s} =$ 14 TeV for p-p collisions (5.5 TeV for Pb-Pb) and to the unprecedent high luminosity,  is expected to open a new era in high energy and heavy-ion physics. 
\begin{table}[h]
\centering
\begin{tabular}{|c||ccc|}
	\hline
\textbf{$\sqrt{s_{NN}}$}  &  Pb-Pb(0-5$\%$) 5.5 TeV    & p-Pb(min. bias) 8.8 TeV  & p-p 14 TeV     \\
	\hline \hline
$\sigma_{NN}^{Q\bar{Q}}$ [mb]  & 4.3 / \textit{0.2} & 7.2 / \textit{0.3} & 11.2 / \textit{0.5}  \\
$N_{tot}^{Q\bar{Q}}$   & 115 / \textit{4.6} & 0.8 / \textit{003} & 0.16 / \textit{0.007}   \\
	\hline
\end{tabular}
\caption{Cross-section and yield for charm and beauty (italic) production obtained from MNR calculations (FO NLO) using EKS98 shadowing. The theoretical uncertainties are of a factor two~\cite{ThM}.}\label{uno}
\end{table}
 It will be an heavy flavour factory where $b\bar{b}$ and $c\bar{c}$  cross sections will increase 100 and 10 times with respect to RHIC. In table \ref{uno} the ALICE baseline for charm and beauty is summarized~\cite{ThM}.
At the LHC energies the Next-to-Leading-Order (NLO) production mechanisms such as gluon splitting and flavour excitation are expected to became competitive with the flavour creation. Moreover, thanks to the wide $\eta$-$p_T$ range covered by the ATLAS, ALICE, CMS and LHCb \cite{lhcb} experiments a new small Bjiorken-x range, down to $10^{-6}$, is available for the investigation. 
Therefore, the detailed study of the heavy flavour production mechanisms represents by itself a strong test of perturbative QCD in an energy regime never investigated before. In addition, the heavy quarks are excellent candidates to probe the Quark-Gluon Plasma (QGP) formed in the nucleus-nucleus collisions. Due to the large virtuality Q, they have a formation time $\Delta t \sim 1/Q \sim 0.1 $ fm/c smaller than the expected formation time of the QGP at LHC ($\sim10$ fm/c).

In addition, less energy loss in the medium is expected for charm and beauty compared to light quarks and gluons due to the suppression of small angle gluon radiation, the so-called \textit{dead-cone effect}~\cite{DeadC}. Therefore, heavy quarks provide the possibility to probe deeper into the hot and dense QCD medium. The use of heavy quark energy loss as a probe to study the QGP properties requires a fine calibration in p-p and p-A, strongly related on the ability to understand the production mechanisms and, in particular, the nuclear initial state effects (e.g effects before the hard scattering) happening in nucleus-nucleus collisions such as $k_T$ broadening and shadowing.  

\section{Quarkonium production and polarization }

\begin{figure}[t]
\begin{minipage}[b]{0.48\linewidth}
\centering
\includegraphics[scale=.47]{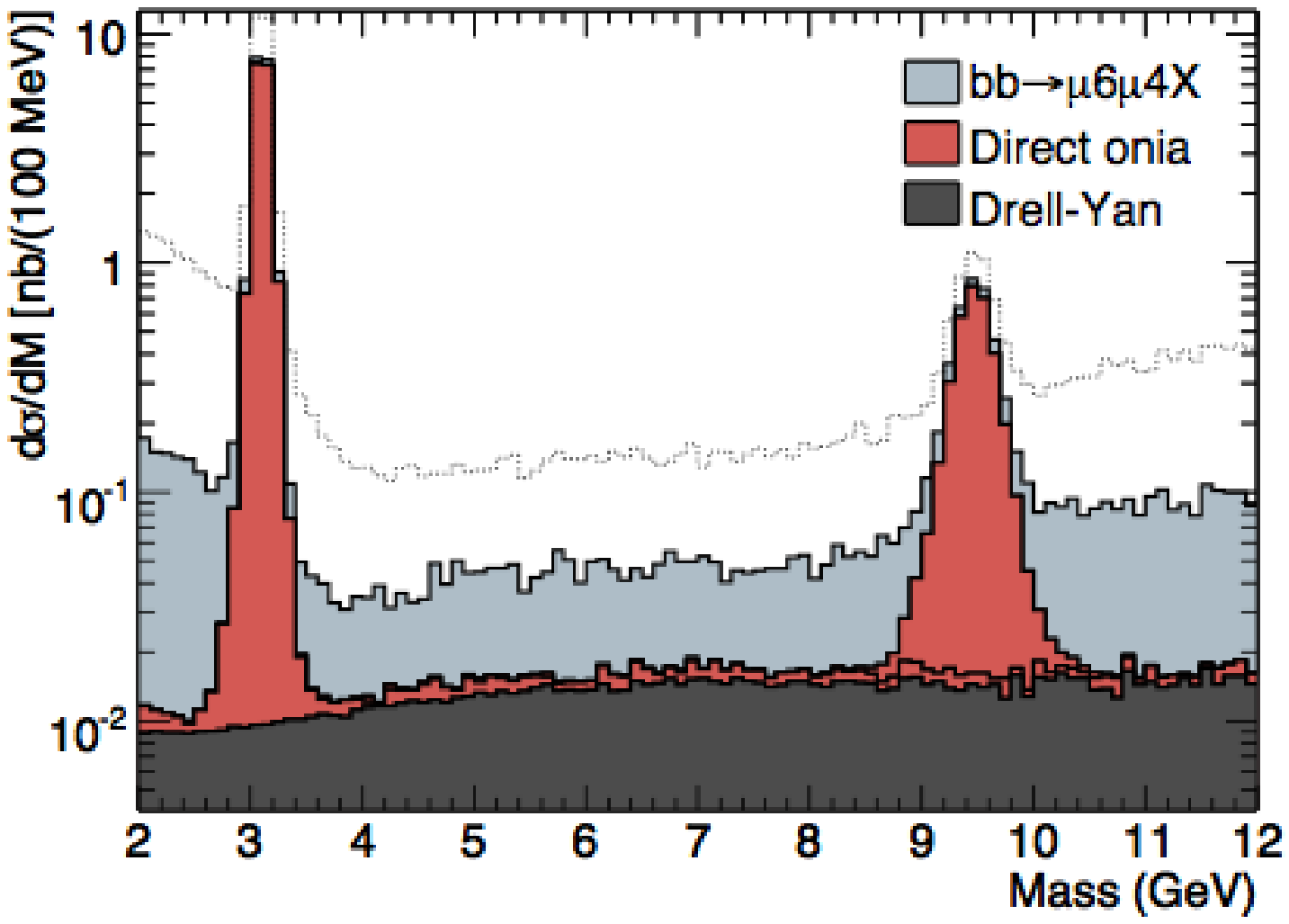}
\caption{Sources of low di-muon invariant mass before (\textit{black dotted line}) and after (\textit{brown}) vertexing and pseudo-proper time cuts.}
\label{fig:tagging2}
\end{minipage}
\hspace{0.4cm}
\begin{minipage}[b]{0.48\linewidth}
\centering
\includegraphics[scale=.60]{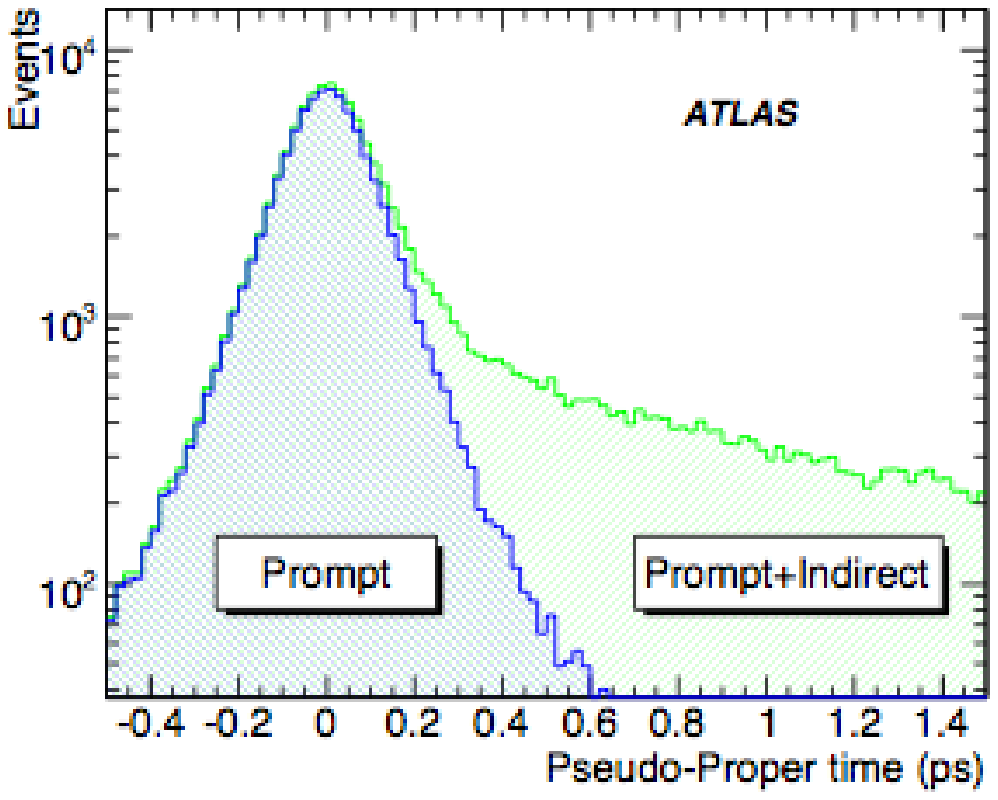}
\caption{Pseudo-proper time distribution for reconstructed prompt $J/\psi$ (\textit{blue shading}) and the sum prompt and indirect (\textit{green shading}).}
\label{fig:tagging3}
\end{minipage}
\end{figure}

Due to high production rate of quarkonia states such as $J/\psi$ and $\Upsilon$ and to their branching fraction into charged leptons some of the first physics results at LHC are expected from quarkonia. From one side these resonances are a good tool for alignment and detector calibration in the early operation phase. 
\begin{figure}[!t]
\begin{minipage}[b]{0.48\linewidth}
\centering
\includegraphics[scale=.65]{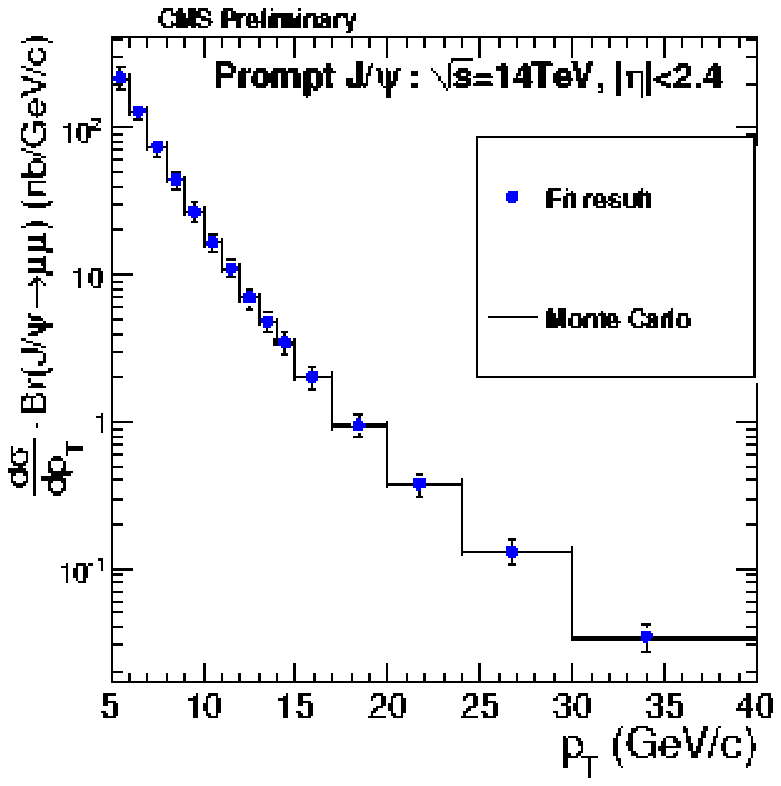}
\caption{Prompt $J/\psi$ differential cross-section versus $p_{T}^{J/\psi}$ integrated over the pseudo-rapidity range $|\eta^{J/\psi}|< 2.4$.}
\label{fig:tagging}
\end{minipage}
\hspace{0.4cm}
\begin{minipage}[b]{0.48\linewidth}
\centering
\includegraphics[scale=.68]{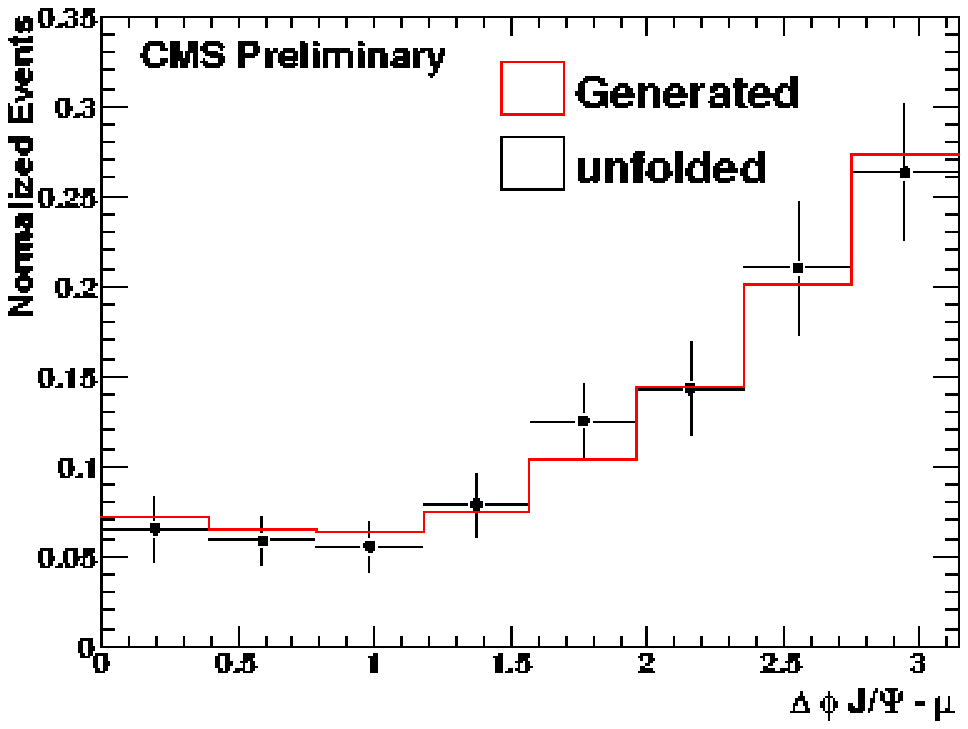}
\caption{$J/\psi$ differential cross section versus the azimuthal angle $\phi$ after unfolding and including systematics uncertainties.}
\label{fig:tagging1}
\end{minipage}
\end{figure}
On the other side the study of the $J/\psi$ and $\Upsilon$ polarization can allow to test QCD predictions and to discriminate between production models even with and integrated luminosity of 10 $pb^{-1}$. Figure~\ref{fig:tagging} illustrates the capability of CMS to study $J/\psi$ and extract the prompt differential cross-section. The result is expected with an integrated luminosity of 3 $pb^{-1}$ \cite{CMSnote1}. 
Going up with the statistic, for an integrated luminosity of 50 $pb^{-1}$ CMS expects to measure the differential cross-section $d\sigma_{b\bar{b}}/d\Delta\phi$ with an initial precision of 15-25$\%$ depending on the $\Delta \phi$ bin (see figure \ref{fig:tagging1})~\cite{cucu}. This analysis is sensible to NLO production mechanisms being the gluon splitting expected to peak at $\Delta \phi$ = 0 while the pair creation peaks at $\Delta \phi$ = $\pi$. In figure \ref{fig:tagging2} the ATLAS approach to $J/\psi$ and $\Upsilon$ studies is presented. Here, the prompt-onia is separated from the \textit{b} decay contribution using the pseudo-proper time distribution as cutting variable (cf. figure \ref{fig:tagging3}).     
In order to extract the $J/\psi$ and $\Upsilon$ polarization~\cite{ATLASref} and discriminate between production models, the data are fitted to the acceptance-corrected $cos\theta^*$ distributions, where $\theta^*$ is the angle between the quarkonium state and the positive lepton. 
With a luminosity of 10 $pb^{-1}$ ATLAS expects to measure the $J/\psi$ polarization with a precision of order of 0.02-0.06, depending on the level of polarization itself in a range of $p_T$ from 10 to 20 GeV/c and beyond. The $\Upsilon$ case is more difficult with first data due to lower cross section and higher background.\\
Moreover, heavy quarkonia measurements are a very promising tool for the investigation of QGP properties. ALICE will use heavy quarkonia studies in p-p collisions as reference for studies in lead-lead collisions. The detector features allow a detailed study down to $p_T$$\sim$ 0 of many channels as $J/\psi$, $\psi'$, $\Upsilon$, $\Upsilon'$, $\Upsilon''$ and $\chi_c$. For reference see \cite{malek, sommer}. 
  
\section{Inclusive $B$ production}

The understanding of inclusive $b$ production and $b$-tagging is a key task not only to test QCD but because $b$-jets represent a source of background for many of the most insteresting physics programs, from Higgs search to top studies. Reference to ALICE, ATLAS and CMS programs can be found in \cite{gue, atlasincl, CMSnote2}.
\begin{figure}[!t]
\begin{minipage}[b]{0.48\linewidth}
\centering
\includegraphics[scale=.6]{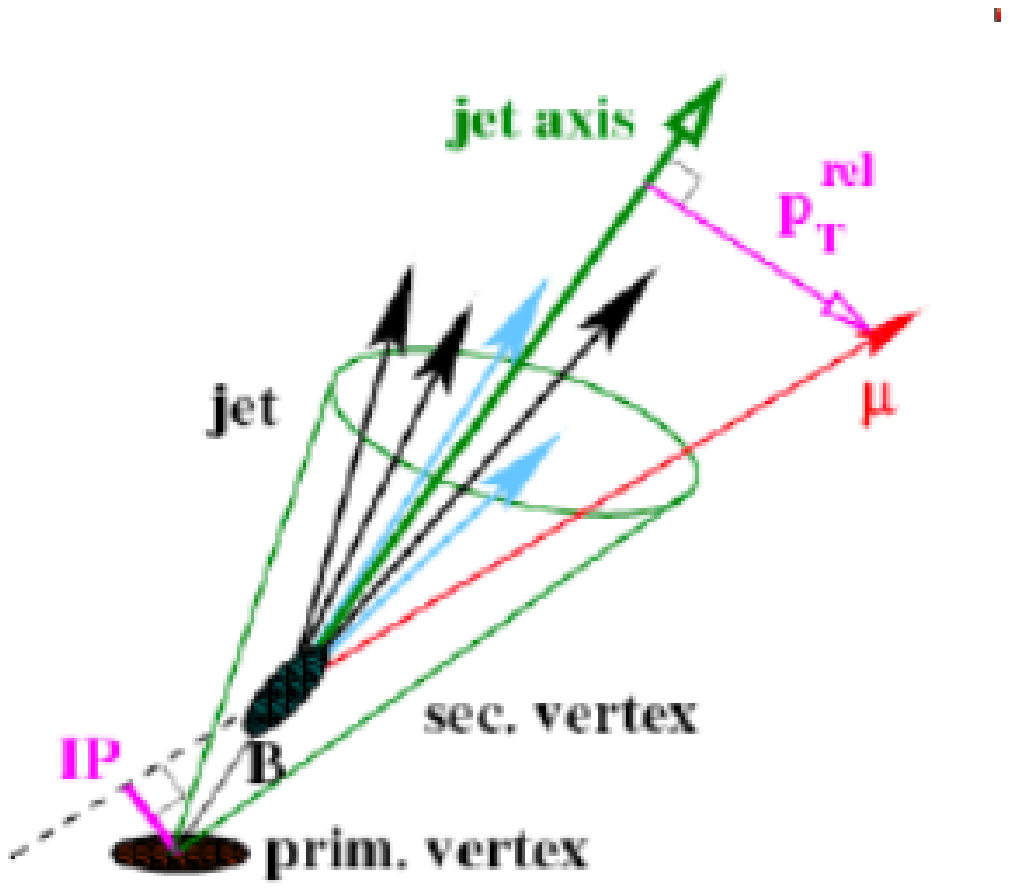}
\caption{Sketch of the b-tagging method in CMS.}
\label{fig:taggi}
\end{minipage}
\hspace{0.4cm}
\begin{minipage}[b]{0.48\linewidth}
\centering
\includegraphics[scale=.58]{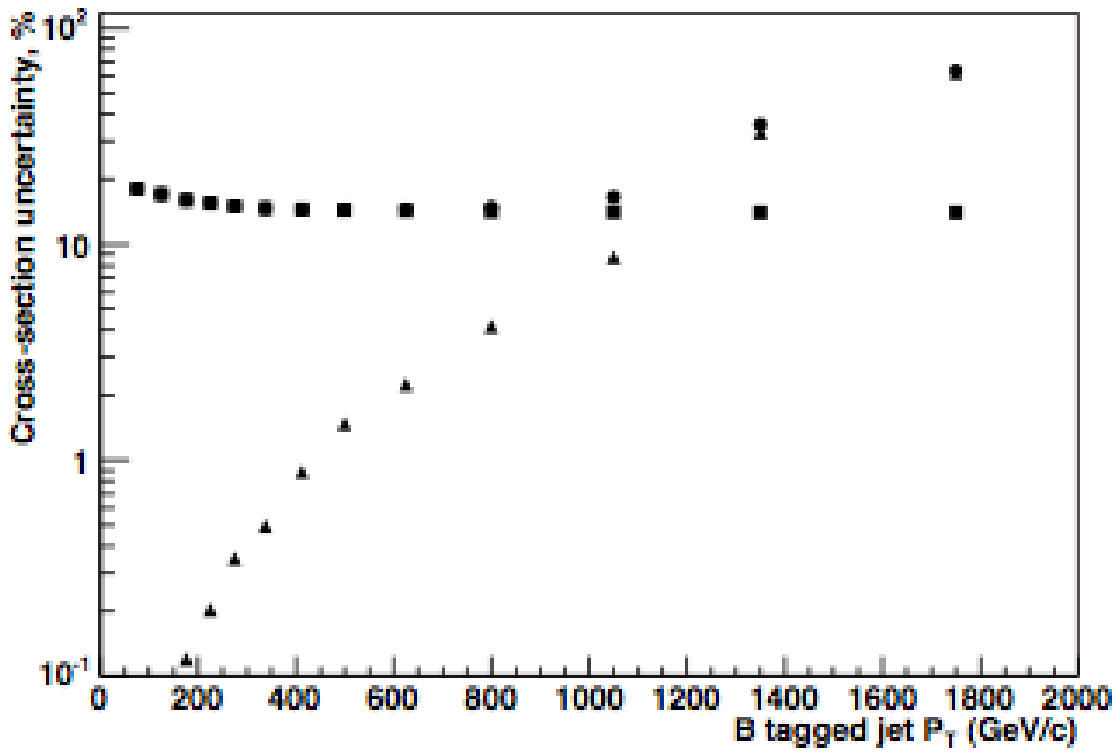}
\caption{Uncertainties on cross-section evaluation.}
\label{fig:taggi2}
\end{minipage}
\end{figure}
 The CMS collaboration carried out a $b$-tagging study in 2006 \cite{CMSrep}, updated at the end of 2008 \cite{CMSnote2}. The trigger is based on a level 1 single muon trigger of 14 GeV/c. The high level trigger request is a single-non isolated $\mu$ with $p_T > 19$ GeV/c and a \textit{b}-tagged jet with $p_T$ $> 50$~GeV/c within the acceptance region of $|\eta|<2.4$. The \textit{b}-tagging algorithm is based on an inclusive secondary vertex reconstruction in jets.
The tagging algorithms combine several topological and kinematic secondary vertex related variables into a single tagging variable to discriminate between jets originating from $b$ quarks and those from light quarks and gluons 
(cf. figure~\ref{fig:taggi}). 
The most energetic $b$-tagged jet in the event is used as $B$-particle candidate to measure the differential cross sections $d\sigma/dp_T$ and $d\sigma/d\eta$. In figure~\ref{fig:taggi2} the estimated statistical and systematic uncertainties are shown. The expected $b$-tagging efficiency is $\sim 65\%$ in the barrel and almost 55$\%$ in the end-cap region.

\section{Exclusive $B$ production}

Production channels such as $B^+ \rightarrow J/\psi K^+$ and $B^0 \rightarrow J/\psi K^{*0}$ represent for CMS and ATLAS good calibration tools and an important reference for future studies of $B_s$ states. CMS expects to measure their differential cross-sections with a statistical precion better than $10\%$ with the first 10~$pb^{-1}$ of statistic \cite{CMSnote3}. With the same statistic ATLAS expect to reach a statistical precision of $\sim5\%$ with systematics of order of 10$\%$ for the $B^+$ channel \cite{ATLASref}. 

\section{Open heavy flavour}

\begin{figure}[t]
\begin{minipage}[b]{0.48\linewidth}
\centering
\includegraphics[scale=.44]{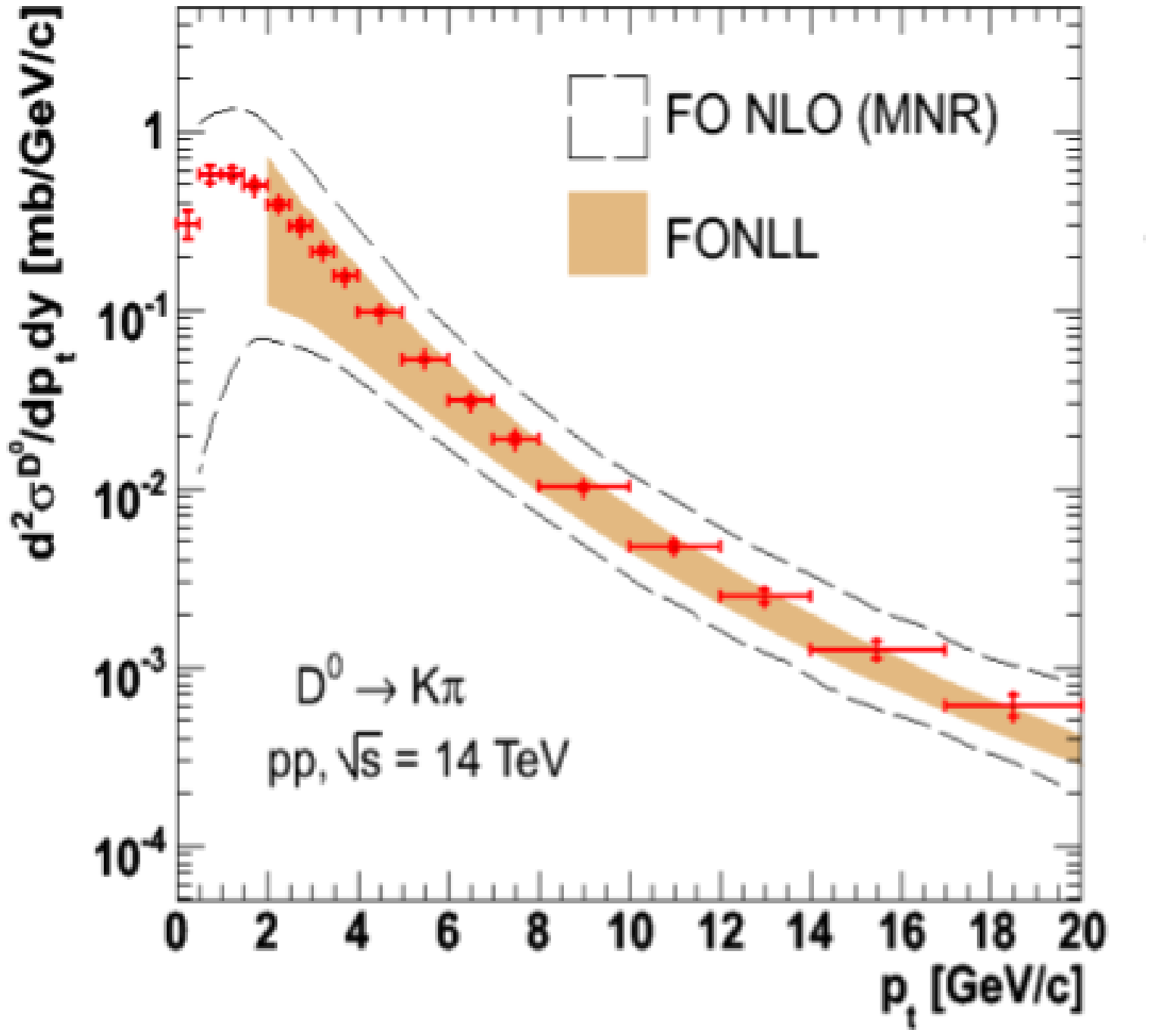}
\caption{Expected $D^{0}$ differential cross-section versus $p_T$.}
\label{fig:tagging7}
\end{minipage}
\hspace{0.2cm}
\begin{minipage}[b]{0.48\linewidth}
\centering
\includegraphics[scale=.44]{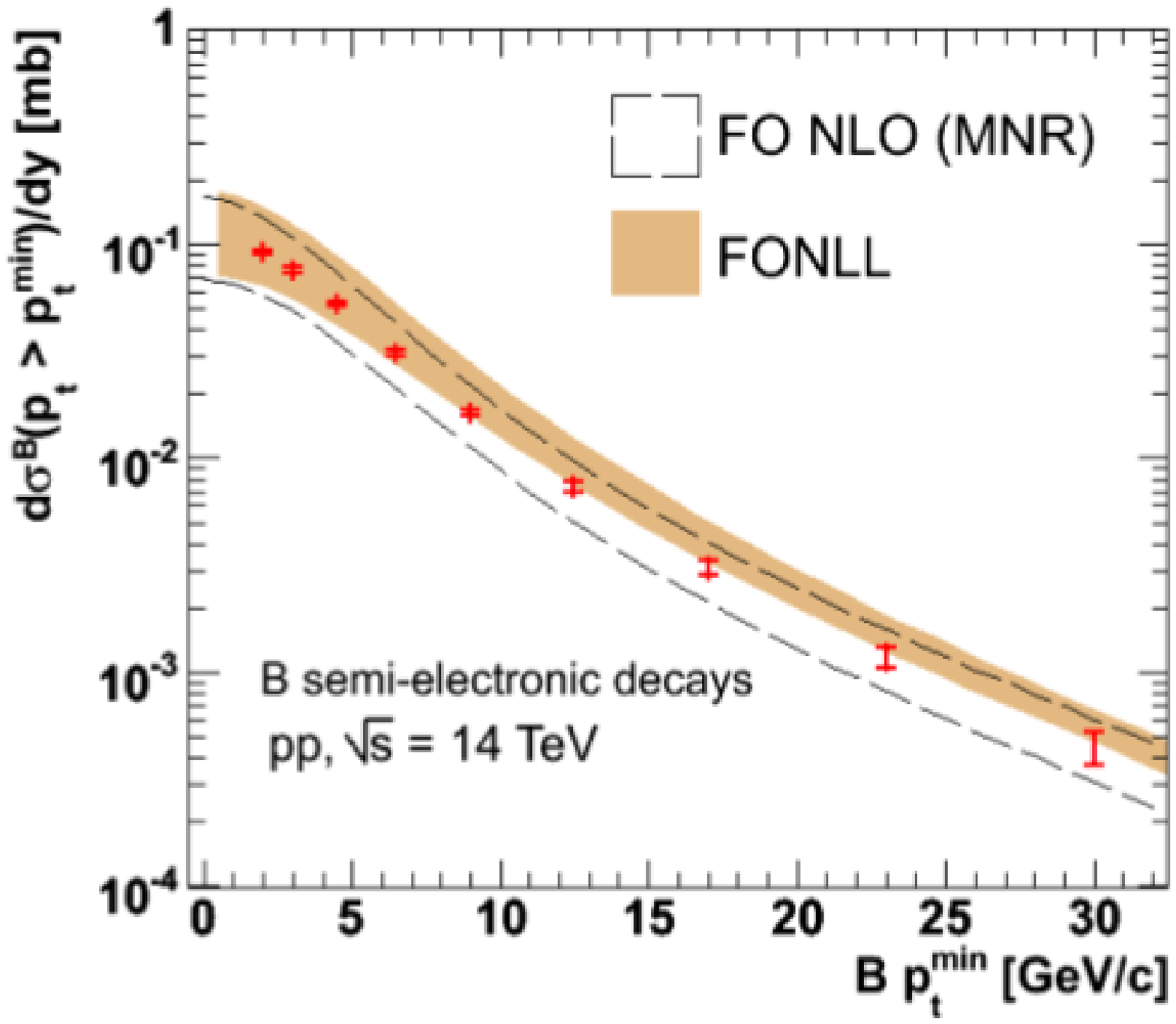}
\caption{Expected differential cross-section versus $p_T$ evaluated for the $b\rightarrow e$ channel.}
\label{fig:tagging8}
\end{minipage}
\end{figure}

The measurement of open charm and beauty production allows to investigate the mechanisms of heavy quark production, propagation and hadronization in the hot and dense medium formed in high energy nucleus nucleus collisions. Moreover, the open charm and beauty cross sections are important for the quarkonia program in ALICE.
\begin{figure}[!t]
\begin{minipage}[b]{0.48\linewidth}
\centering
\includegraphics[scale=.5]{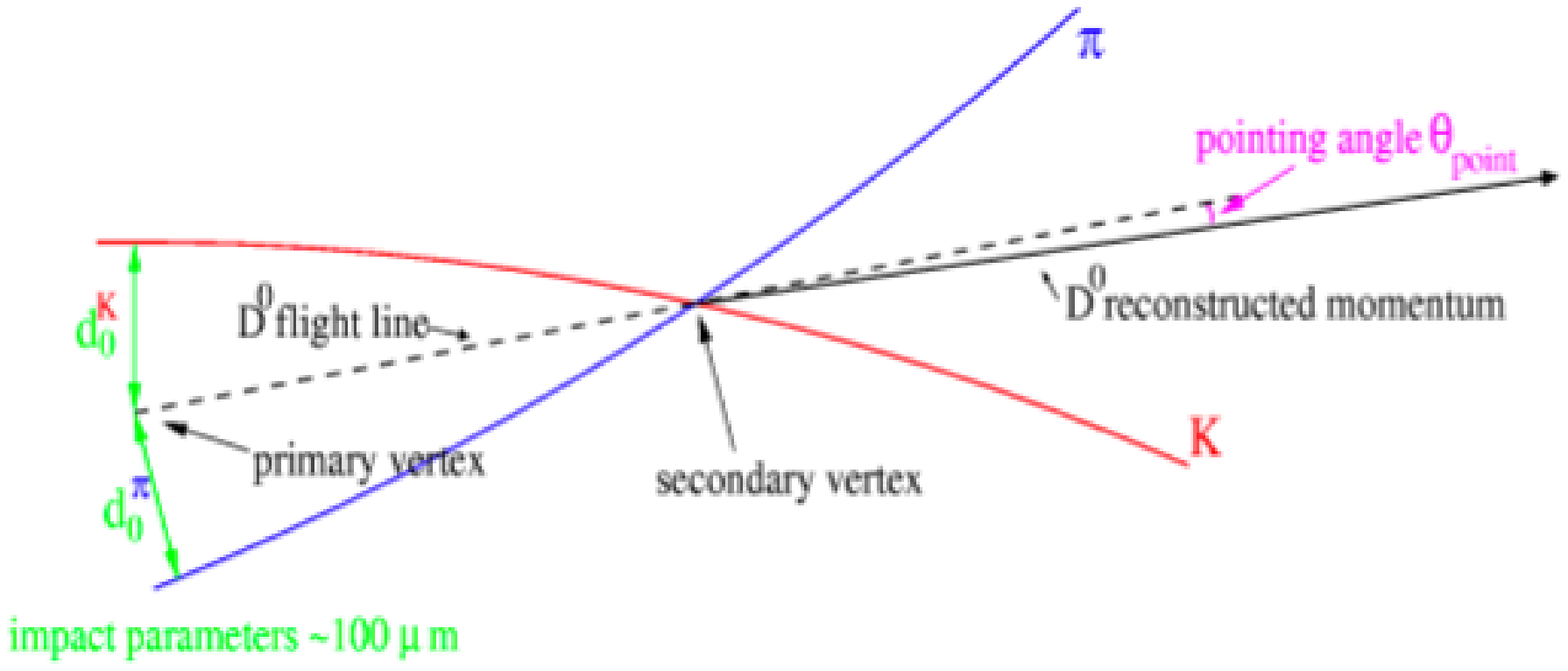}
\caption{Sketch of the $D^0\rightarrow K\pi$ decay topology considered in ALICE.}
\label{fig:tagging5}
\end{minipage}
\hspace{0.4cm}
\begin{minipage}[b]{0.48\linewidth}
\centering
\includegraphics[scale=.5]{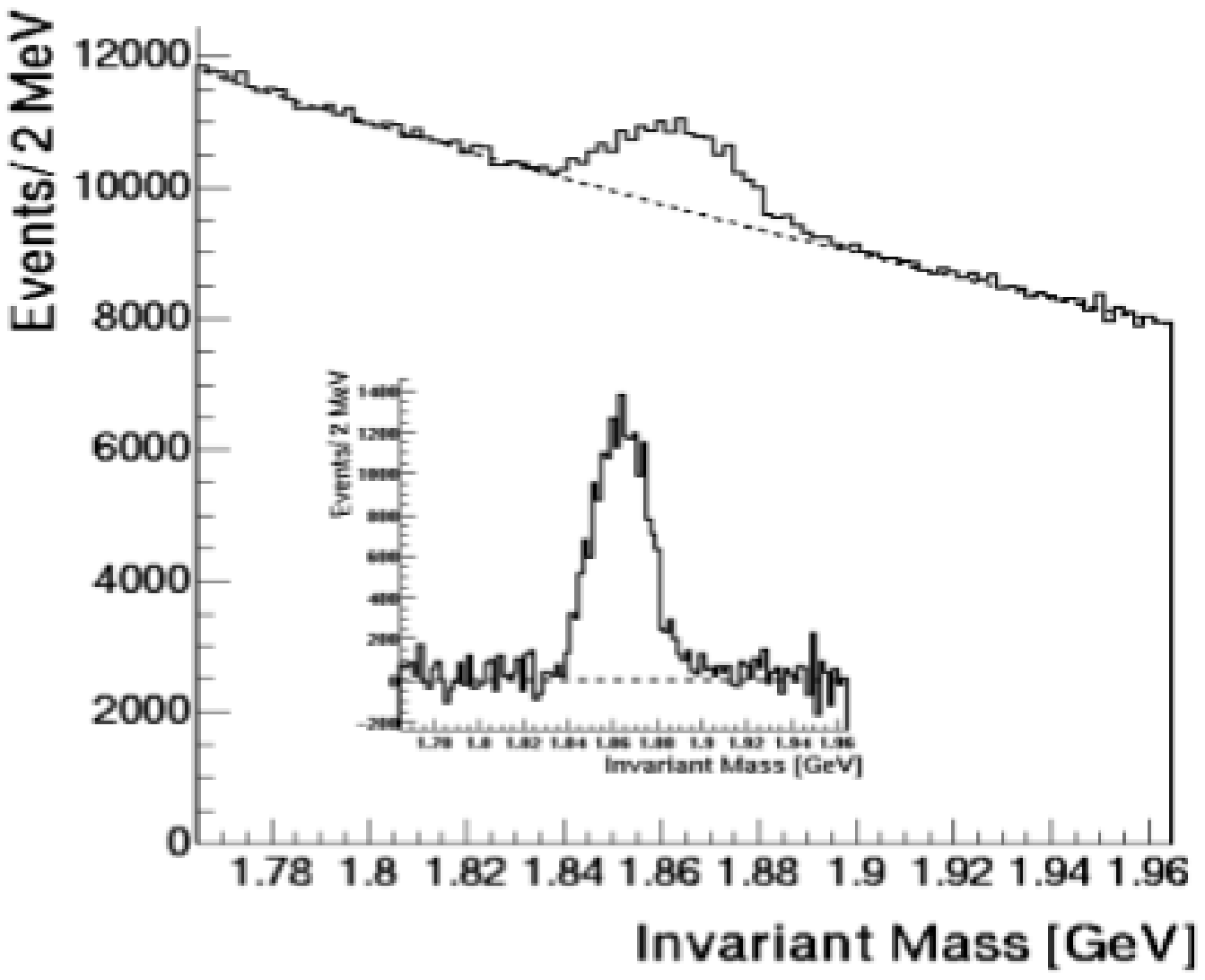}
\caption{$K\pi$ invariant mass distribution before and after background subtraction (inset) expected for one year of Pb-Pb data taking \cite{dainese}.}
\label{fig:tagging6}
\end{minipage}
\end{figure} 
The ALICE detector was designed and optimized to study Pb-Pb collisions at $\sqrt{s_{NN}} = 5.5$ TeV, providing excellent capabilities for tracking and particle identification. The strategy for the exclusive reconstruction of $D^0 \rightarrow K\pi$ (branching ratio $\sim 3.86\%$), key channel for open charm studies, is based on:
\begin{itemize}
\item Displaced vertex topologies. Tracks with large impact parameter and small pointing angle are selected. The decay topology is illustrated in figure \ref{fig:tagging5}.
\item Identification of the kaon using the TOF detector.
\item Invariant mass analysis as illustrated in figure~\ref{fig:tagging6}.
\end{itemize}
The $D^{0}$ selection cuts are optimized separately for p-p, p-A and A-A collisions. The $D^0$ $p_T$ spectrum for one year of p-p data taking is depicted in figure~\ref{fig:tagging7}. The statistical error is expected to be in the order of 15-20$\%$ and the systematic uncertainties less than 20$\%$ \cite{dainese}.

Another promising channel for heavy-flavour studies is the semi-leptonic decay of open beauty ($B\rightarrow e + X$) \cite{ALICEnote1}. Here, the detection strategy is based on three points. At first order the electron particle identification using TPC and TRD detectors is expected to reject most of the hadrons. The background from charm, the contribution from Dalitz decays and $\gamma$ conversions are suppressed cutting on the small impact parameter values $< 200$ $ \mu m$.  
The small residual background will be removed using ALICE data and Monte Carlo simulations. ALICE has many other channels for \textit{D} and \textit{B} under study to better constrain the production cross-section. The expected precision, equivalent to one year of data taking at the nominal ALICE working luminosity of $\mathcal{L}$ = $3 \times 10^{30} cm^{-2}s^{-1}$ is shown in figure~\ref{fig:tagging8}.
To improve the knowledge about the different charm production mechanisms, an analysis to measure the gluon splitting rate is in development \cite{agrelli}. Here, the $D^{*\pm}$ content in jets is investigated, and using the different fragmentation charateristic of gluon splitting and pair creation the two contribution can be separated.

\section{Conclusions}

We reviewed parts of the ALICE, ATLAS and CMS heavy-flavour programs. In the early phase of data taking a focus will be in the understanding of the different production mechanisms. On one side these studies are a powerful tool to test QCD and on the other side they are the baseline for $B_s$ studies and new physics search in ATLAS and CMS and the reference for heavy flavour measurements in heavy-ion collisions in ALICE.

\end{document}